\documentclass[twocolumn,prl,superscriptaddress,showpacs]{revtex4}
\usepackage{amsmath,amssymb}
\usepackage{graphicx}
\usepackage{verbatim}
\usepackage{amsfonts}
\usepackage{dcolumn}
\usepackage{bm}
\usepackage{color}
\usepackage[colorlinks,citecolor=blue]{hyperref}

\begin{document}

\title{\large \bf  Symmetry-Protected Quantum Spin Hall phases in Two Dimensions}

\author{Zheng-Xin Liu}
\affiliation{Institute for Advanced Study, Tsinghua University, Beijing, 100084, Peoples's Republic of China}
\affiliation{Department of Physics, Massachusetts Institute of Technology, Cambridge, Massachusetts 02139, USA}
\author{Xiao-Gang Wen}
\affiliation{Department of Physics, Massachusetts Institute of Technology, Cambridge, Massachusetts 02139, USA}
\affiliation{Perimeter Institute for Theoretical Physics, Waterloo, Ontario, N2L 2Y5 Canada}

\begin{abstract}
Symmetry-protected topological (SPT) states are short-range entangled states
with symmetry.  Nontrivial SPT states have symmetry protected gapless edge
excitations.  In 2 dimension, there are infinite number of nontrivial SPT
phases with $SU(2)$ or $SO(3)$ symmetry. These phases can be described by
$SU(2)$ or $SO(3)$ nonlinear-sigma models with a \emph{quantized} topological
$\theta$ term. At an open boundary, the $\theta$ term becomes the
Wess-Zumino-Witten term and consequently the boundary excitations are decoupled
gapless left movers and right movers. Only the left movers (if $\theta>0$)
carry the $SU(2)$ or $SO(3)$ quantum numbers.  As a result, the $SU(2)$ SPT phases
have a half-integer quantized spin Hall conductance and the $SO(3)$ SPT phases
have an even-integer quantized spin Hall conductance.  Both the  $SU(2)$ and $SO(3)$ SPT
phases are symmetric under their $U(1)$ subgroup and can be viewed as $U(1)$
SPT phases with even-integer quantized Hall conductance.

\end{abstract}


\maketitle

Gapped quantum states may belong to long-range entangled phases or short-range
entangled (SRE) phases \cite{CGW1038}. Long-range entangled states have intrinsic
topological order and cannot be deformed into direct product states through
finite steps of local unitary transformations. Examples of intrinsic
topologically ordered phases include fractional quantum Hall
liquids\cite{TSG8259,L8395}, chiral spin liquids\cite{KL8795,WWZ8913}, and
$Z_2$ spin liquid \cite{RS9173,W9164,MS0181}. On the other hand, SRE states are
equivalent to direct product states under local unitary transformations. If
there is no symmetry, there will be only one SRE phase. If the system has a
symmetry, the phase diagram will be much richer.  Even SRE states which do not
break any symmetry can belong to different phases.  Those phases are called SPT
phases which stands for symmetry-protected topological phases or symmetry-protected trivial phases. The well-known Haldane phase in $S=1$ spin
chain\cite{H8364,AKL8877} is the first example of bosonic SPT phase in 1 dimension(1D),
which is protected by either $D_2$ spin rotation symmetry or time reversal
symmetry.  Topological insulators \cite{KM0502,BZ0602,MB0706,FKM0703,QHZ0824}
are 2 dimension (2D) SPT phases in free fermion systems protected by time reversal symmetry
$T$ and $U(1)$ charge conservation symmetry.

Some thought that the topological insulators are characterized by
quantum spin Hall effect. 
However, since spin rotation symmetry is broken by spin-orbital coupling, spin angular momentum is not conserved.
Therefore, there is no spin Hall effect in usual topological insulators.
Quantum spin Hall effect will be present only if the topological insulators
also have an extra $U(1)$ spin rotation symmetry\cite{KM0501}.

In this Letter, we will introduce another kind of SPT phases ---
$SU(2)$ or $SO(3)$ SPT phases in 2D, which are classified by $\mathbb Z$.
In contrast to topological insulators, these phases are interacting bosonic phases.
Owning to the $SU(2)$ or $SO(3)$ symmetry, if the system is open, the boundary
excitations will be gapless although the bulk remains gapped. Importantly, different
SPT phases can be distinguished experimentally through their linear responses.
To this end, we couple the model to external probe field, which is an analogue
of the electromagnetic field for spins. We show that spin Hall current will be
induced on the boundary with a quantized spin Hall conductance.  Different
$SU(2)$ SPT phases are characterized by their different half-integer quantized
spin Hall conductance, while different $SO(3)$ SPT phases by even-integer
quantized spin Hall conductance.

\textit{$SU(2)$ principal chiral Nonlinear Sigma Model (NLSM)}. ---
In 2D, $SU(2)$ SPT phases are classified by group cohomology class $\mathcal
H^3(SU(2),U(1))=\mathbb Z$ \cite{CGL1172}.  Owning to the
correspondence between the group cohomology class
and the topological cohomology class \cite{DW9093},
each SPT phase can be described by a principal chiral NLSM with quantized topological
$\theta$ term [which is classified by  $H^3(SU(2),Z)=\mathbb Z$].  The $\theta$ term of
the NLSM can be written as\cite{Xu}
\begin{eqnarray}\label{Topterm}
S_{\mathrm{top}}=-i\frac{\theta}{24\pi^2}\int_{M} \mathrm{Tr} (g^{-1}dg)^3,\ \ \ g\in SU(2)
\end{eqnarray}
where $M$ is the Euclidian space-time manifold,  $g \in SU(2)$ is a
$2\times 2$-matrix-valued function of
space-time $g(x)$, and $\theta=2\pi K$ with $K\in\mathbb Z$ corresponding to the $K$th $SU(2)$ SPT phase.
When $M$ has no boundary, $S_{\mathrm{top}}$ is
quantized into integer times of $-2\pi i$.

Including the dynamic part, the partition function of the NLSM is $Z=\int \mathcal Dg e^{-\int_M d^3x \mathcal L}$, where $\mathcal L$ is the
Lagrangian density,
\begin{eqnarray}\label{S}
\mathcal L&=&-{1\over4\lambda^2}\mathrm{Tr}[(g^{-1}\partial_\mu g)(g^{-1}\partial_\mu g)]\nonumber\\
&&-i{K\over12\pi}\mathrm{Tr}(\varepsilon^{\mu\nu\gamma}g^{-1}\partial_\mu gg^{-1}\partial_\nu gg^{-1}\partial_\gamma g).
\end{eqnarray}
For large enough $\lambda^2$, the renormalization flows to a fixed point where only
the topological term remains ($\lambda^2$ flows to infinity). The fixed point Lagrangian
captures all the physical properties of the SPT phases. So we will focus
on the fixed point in the following discussion.

Since the symmetry group is of crucial importance for the physical properties of
SPT phases, we stress that the symmetry group of our system is $SU(2)_L$, under which
the group element $g$ varies as $g\to \hat h g=hg$ for $\hat h\in SU(2)_L$. It is easy to
check that the Lagrangian equation~(\ref{S}) is invariant under $SU(2)_L$. It can be shown that
equation~(\ref{S}) has a larger symmetry, it is invariant under the group $SU(2)_L\times SU(2)_R$, where
$SU(2)_R$ is the right multiplying group defined as $\hat{\bar
h}g= gh^{-1},\ \hat{\bar h}\in SU(2)_R$. Furthermore, equation~(\ref{S})
also has time reversal symmetry $T$.  Namely, it is invariant under the time
reversal transformation, $t\to -t$, $i\to -i$ (consequently $\tau\to
\tau$), and $g\to g^{-1}$ \cite{T}. The SPT phases only need the protection of $SU(2)_L$.
As will be discussed later, when the extra symmetry $SU(2)_R$ and $T$ is removed by perturbation
$\delta {\cal L}=\mathrm{Tr}[\partial_\tau gM(x)g^{-1}]$ with $M(x)$ external field, the physical
properties of the SPT phases remains unchanged. In the following we will discuss
the Lagrangian equation~(\ref{S}) and note $SU(2)_L$ as $SU(2)$ without causing confusion.

If the system has a boundary,  the quantized $\theta$ term equation~(\ref{Topterm})
becomes the Wess-Zumino-Witten term\cite{WZ7195,W8322} in the 1+1D
boundary effective theory. According to Ref.~\onlinecite{W8455}, a 1+1D Wess-Zumino-Witten
model with given $K$ may flow to a gapless fixed point
$S_{\mathrm{bdr,fix}}={|K|\over8\pi}\int dx^0
dx^1\mathrm{Tr}[(g^{-1}\partial_\mu g)(g^{-1}\partial_\mu
g)]+S_{\mathrm{top}}$, where $S_{\mathrm{top}}$ are defined in equation  ~(\ref{Topterm}),
$x^0=\tau$ is the imaginary time and $x^1$ is the spacial dimension along the boundary.

If $K>0$, the boundary excitations at the fixed point are decoupled left mover
$J_+= {K\over2\pi} \partial_+gg^{-1}$ and right mover $J_-=-{K\over2\pi}g^{-1}\partial_-g$,
where $x^\pm={\sqrt{1\over2}}(x^0\pm ix^1)$ is the chiral coordinate and $\partial_\pm={\sqrt{1\over2}} (\partial_0
\mp i\partial_1)$.  $J_\pm$ satisfy the
equation of motion $\partial_\mp J_\pm=0$ (which yields gapless dispersion). Importantly, $J_+$ and $J_-$ behave
differently under global $SU(2)$ transformation $g\to hg$. The current $J_-$
is $SU(2)$ invariant $J_-\to J_-$, but $J_+$ is $SU(2)$ covariant $J_+\to
hJ_+h^{-1}$, \textit{so only the left mover $J_+$ carries $SU(2)$ ``charge''.}
This property indicates that the gapless boundary excitations are protected by
the $SU(2)_L$ symmetry, because the mass term, such as ${\cal L}_{\rm{bdr,mass}}
\propto(\mathrm{Tr}g)^2$ \cite{mass}, which gaps out the excitations
will mix the left mover and right mover and hence breaks the $SU(2)_L$ symmetry.
The bulk perturbation $\delta {\cal L}=\mathrm{Tr}[\partial_\tau gM(x)g^{-1}]$, on the other hand,  will not cause
scattering between the left mover and the right mover since it respects $SU(2)_L$ symmetry;
hence, it will leave the boundary excitations gapless.
Under time reversal $T$, $J_+$ and $J_-$ exchange their roles $J_+
\leftrightarrow J_-$. If $K<0$, then the boundary excitations will be redefined as
$J_+=-{K\over2\pi}g^{-1}\partial_+g,\ \ J_-= {K\over2\pi} \partial_-gg^{-1}$.
In this case, $J_+$ is $SU(2)$ neutral and $J_-$ carries $SU(2)$ charge.


Following Ref.~\onlinecite{W8455}, the boundary excitations of the $SU(2)$ SPT state labeled by $K$ are described by $SU(2)$ Kac-Moody algebra of level $|K|$. In the following we will study how the system (especially the boundary) responds to an external probe field. Without loss of generality, we assume $K>0$.

\textit{Quantized spin Hall conductance}. ---
Now we introduce an external probe field $A$, which minimally couples to the
topological NLSM by replacing every $g^{-1}\partial_\mu g$ term with
$g^{-1}(\partial_\mu+A_\mu)g$. Expanding $A$ by three Pauli matrices,
$A={1\over2}\sum_{\mu,a}A_\mu^a\sigma^ad x^\mu$, then we can define a current
density operator $J_\mu^a={\delta\mathcal L\over\delta A_\mu^a}|_{A_\mu=0}$
with
\begin{eqnarray*}\label{current}
J^a_\mu = -{1\over2\lambda^2}\mathrm{Tr}(\partial_\mu gg^{-1} {\sigma^a\over2} )+ i{K \over4\pi}\varepsilon_{\mu\nu\gamma}  \partial_\gamma\left[{\mathrm{Tr}}(\partial_\nu g g^{-1}{\sigma^a\over2})\right].
\end{eqnarray*}
$J_\mu^a$ is the conserved spin current corresponding to the global $SU(2)$
invariance of the action. The second term on the right-hand side contributes a boundary current since
it is a total differential.

At the fixed point $\lambda^2\to\infty$, only the topological term remains,
\begin{eqnarray}\label{A}
-i{K\over12\pi}\mathrm{Tr}[&g^{-1}&(d+A)g]^3
=-i{K\over12\pi}\mathrm{Tr} [(g^{-1}dg)^3+A^3
\nonumber\\
&&+3(dgg^{-1}\wedge F)+ 3d(dgg^{-1}\wedge A)].
\end{eqnarray}
Notice that equation~(\ref{A}) is invariant under local $SU(2)$ transformation $g\to
hg$, if the field $A$ varies as $A\to hAh^{-1}+hdh^{-1}$. If $F=0$, then $A$
only couples to the edge current via $\mathrm{Tr} (dgg^{-1}\wedge A)$. Notice
that only the right moving component $J_+$ occurs in $dgg^{-1}$. This means
that $A$ only couples to $J_+$ and does not couple to $J_-$. When $F\neq0$, the
bulk term $3\mathrm{Tr} (dgg^{-1}\wedge F)$ in equation~(\ref{A}) is difficult to treat.
In order to obtain an effective field theory of the external field $A$ and $F$,
we need to integrate out the group variables $g$.

To avoid this difficulty, we take the advantage of the local ``gauge invariance''
of the Lagrangian in equation~(\ref{A}). Here the local ``gauge transformation'' is defined
as $g\to h(x)g$ and $A\to hAh^{-1}+hdh^{-1}$. When integrating out the group
variables, the effective action of $A$ should also be ``gauge'' invariant. So we
expect the result is the Chern-Simons action (we will see later that this
effective action is self-consistent),
\begin{eqnarray*}\label{SA}
S_{\mathrm{eff}}(A)&=& i{K\over4\pi}\int_M \mathrm{Tr}(A\wedge F-{1\over3}A^3),
\nonumber\\&=& i{K\over8\pi}\int_M d^3x\varepsilon^{\mu\nu\lambda}(A^a_\mu\partial_\nu A^a_\lambda+\varepsilon_{abc}{i\over3} A^a_\mu A^b_\nu A^c_\lambda ),\nonumber
\end{eqnarray*}
where $A=\sum_a A^a_\mu dx^\mu {\sigma_a\over2} $.  Notice that the trace
$\mathrm {Tr}({\sigma_a\over2}{\sigma_b\over2})={1\over2}\delta_{ab}$
contributes an extra coefficient $1\over2$. If $F=0$,
$S_{\mathrm{eff}}(A)=-i{K\over12\pi}\int \mathrm{Tr}A^3$, which is consistent
with equation~(\ref{A}). From the above effective action, we obtain the response current
density,
\begin{eqnarray}\label{respons}
\mathcal J_\mu^a={\delta S_{\mathrm{eff}}\over\delta A_\mu^a}= i{K\over4\pi }\varepsilon_{\mu\nu\lambda} (\partial_\nu A_\lambda^a+{i\over2} \varepsilon_{abc} A_\nu^bA_\lambda^c).
\end{eqnarray}

It will be easier to see the response of the system if the probe field $A$ only
contains the spin-$z$ component, $A=\sum_\mu A^z_\mu {\sigma_z\over2}dx^\mu$, which
can be viewed as the spin-electromagnetic field that couples to $S^z$ as
its charge.  Then the responding spin density is proportional to the
``magnetic field'',
\[\mathcal J_0^z=
i{K\over4\pi}(\partial_1 A_2^z-\partial_2A_1^z)= i{K\over4\pi}b^z.
\]
Here we use $0,1,2$ to label the space-time index and $x,y,z$ to label the spin
direction. The spin current is proportional to the ``electric field'',
\begin{eqnarray*}
\mathcal J_1^z&=& i{K\over4\pi}(\partial_2A_0^z-\partial_0A_2^z)={K\over4\pi}e_2^z,\\
\mathcal J_2^z&=& i{K\over4\pi}(\partial_0A_1^z-\partial_1A_0^z)=-{K\over4\pi}e_1^z.
\end{eqnarray*}
The direction of the motion of the spin current is orthogonal to the direction
of the ``electric field''. This is nothing but a spin Hall effect. Furthermore,
the spin Hall conductance is quantized as ${K\over4\pi}$, which is half of the
electric integer charge Hall conductance. From this information,
we conclude that the $SU(2)$ symmetric topological NLSM
model describes a bosonic spin quantum Hall system.

The $SU(2)$ SPT phases can also be viewed as $U(1)$ SPT phases, where $U(1)$ is
the $S^z$ spin rotation.  The above result implies that the $U(1)$ SPT phases are
characterized by a quantized Hall conductance.  To understand the value of
quantization, let us introduce $ A^c_\mu =\frac12  A^z_\mu$.  The charge that
$A^c_\mu$ couples to is $2S^z$ which is quantized as integers.
The effective action for $ A^c_\mu$ is given by
$S_{\mathrm{eff}}(A^c)=
i{K\over2\pi}\int_M d^3x\varepsilon^{\mu\nu\lambda}
A^c_\mu\partial_\nu A^c_\lambda.$
We see that the charge Hall conductance is $\frac{2K}{2\pi}$.  In other words,
the Hall conductance for the $U(1)$ SPT phases is quantized as \emph{even}
integers $2K$ (in unit of $\frac{1}{2\pi}$), which agrees with a calculation by
$U(1)\times U(1)$ Chern-Simons theory \cite{L,LV1256}.

In the electric integer quantum Hall system, the boundary excitations are chiral
currents. In contrast, the boundary of model (\ref{Topterm}) contains both
left-moving and right-moving gapless excitations. However, only the left mover carries
$SU(2)$ charge and couples to the probe field $A$. In other words, the $A$
field will induce left-moving spin current.

The coupling of the left moving current to $A$ field is consistent with the
Chern-Simons action. Remembering that the topological term (\ref{A}) is local
`gauge invariant'.  If space-time is closed, the effective action (\ref{SA}) is
gauge invariant as expected. However, if space-time has a boundary, equation~(\ref{SA})
is no longer gauge invariant. Under local gauge transformation $A\to
A'=hAh^{-1}+hdh^{-1}$, the variance of the Chern-Simons term is
\begin{eqnarray}\label{Anm_CS}
S_{\mathrm{eff}}(A')-S_{\mathrm{eff}}(A)&=& i{K\over4\pi}\left[\int_{\partial M} \mathrm{Tr}(h^{-1}dh\wedge A)\right.\nonumber\\&&\left.+\int_M {1\over3}\mathrm{Tr}(h^{-1}dh)^3\right].
\end{eqnarray}
The first term on the right hand side depends on the values of $A$ on the
boundary, and the second term is independent on $A$.

Since the gauge anomaly in equation~(\ref{Anm_CS}) is purely a boundary term, it can be
canceled by a matter field on the boundary described by
$SU(2)$ level-$|K|$ Kac-Moody algebra.  To see
the cancelation of the anomaly, we may
embed the $SU(2)$ level-$|K|$ Kac-Moody algebra into $|K|$ spin-1/2 complex
fermions $\psi_I, (I=1,2,...,K)$, which leads to the following effective edge theory:
\begin{eqnarray*}\label{Boundary}
S_{\mathrm{bdr}}(\psi,A)&=&  \int dx^0 dx^1 \sum_{I=1}^K\left[\right.\psi^\dag_{I-}(\partial_0-i\partial_1) \psi_{I-}\nonumber\\&&\left.+\psi^\dag_{I+} [(\partial_0+{}A_0)+i(\partial_1+{}A_1)]\psi_{I+}\right].
\end{eqnarray*}
Under gauge transformation $\psi_+'=h\psi_+$, $A'=hAh^{-1}+hdh^{-1}$, the above action has an anomaly\cite{ZWZ8477,AG8449}(for details, see the Supplemental Material\cite{supmat})
$S_{\mathrm{bdreff}}(A')-S_{\mathrm{bdreff}}(A)=-i{K\over4\pi}\int_{\partial M} \mathrm{Tr}(h^{-1}dh\wedge A),$
which exactly cancels the anomaly of the Chern-Simons action in equation~(\ref{Anm_CS}).
This means that the total action of bulk Chern-Simons term and the boundary
fermion term is gauge invariant (up to a term which is independent on $A$).

Since we have $K$ flavors of fermion fields, they also form a representation of
$U(K)$ Kac-Moody algebra, which gives rise to extra gapless edge modes.
However, only the representation of $SU(2)_K$-Kac Moody algebra are physical
degrees of freedom in our model. The extra gapless modes can be gapped out by
mass terms which do not break the $SU(2)$ symmetry, or can be removed by
performing a projection onto the $U(k)$ singlet at each site \cite{W9125}.

Supposing $\bar A$ is the time reversal partner of $A$, then under $T$
transformation, $\partial_\tau\to \partial_\tau$, $i\to -i$, $g\to g^{-1}$, $A_\mu\to  \bar
A_\mu$, the Lagrangian (\ref{A}) becomes 
\begin{eqnarray*}
i{K\over12\pi}\mathrm{Tr}[g(d&+&\bar A)g^{-1}]^3=-i{K\over12\pi}\mathrm{Tr}[(g^{-1}dg)^3-\bar A^3\\&&+3(g^{-1}dg\wedge\bar F)+3d(g^{-1}dg\wedge \bar A)]
\end{eqnarray*}
where $\bar F=d\bar A+\bar A\wedge\bar A$. From the above equation, we can see that
$\bar A$ only couples to $J_-$, which carries $SU(2)_R$ charge and is $SU(2)_L$
neutral. Thus the time reversal operation $T$ transforms the $SU(2)_L$
quantities $A$ and $J_+$ to the $SU(2)_R$ quantities $\bar A$ and $J_-$. This
is very different from the model with $-K$, where the right mover $J_-$ carries
$SU(2)_L$ charge and is coupling to $A$.

\textit{$SO(3)$ SPT phase in 2+1D}. --- Above we discussed a bosonic spin-1/2 model
with quantized spin Hall effect. However, a bosonic particle can never carry
spin-1/2. So the $SU(2)$ SPT phases only have theoretical interest. In the
following, we will discuss a more realistic bosonic model of integer spins,
whose symmetry group is $SO(3)$,
\begin{eqnarray}\label{TopSO3}
S_{\mathrm{top}}=-i\frac{2\pi K}{2\times48\pi^2}\int_{M} \mathrm{Tr} (g^{-1}dg)^3, \ \ \ g\in SO(3).
\end{eqnarray}
Here $g\in SO(3)$ is a $3\times3$ matrix, and $K\in \mathbb Z$ is an element of the cohomology
$H^3(SO(3),Z)=\mathbb Z$ which is generated by $\frac{1}{48\pi^2}\int_{G} \mathrm{Tr}
(g^{-1}dg)^3$. The factor 2 in the denominator of equation~(\ref{TopSO3}) is owing to
the factor that closed space-time manifold (e.g. $M=S^3$) must cover the group manifold
$G=S^3/Z_2$ even times.

Above topological action (\ref{TopSO3}) should be quantized to integer times of $-2\pi
i$, even if $M$ is the group manifold itself. To satisfy this condition,
$K$ must be an even integer. In other words, only even $K$ belongs to $H^3(SO(3),Z)=\mathbb Z$.
Furthermore, only $K = 4r$, $r\in \mathbb{Z}$ give rise to SPT phases. The
mathematical reason is that the map from the group cohomology
$\mathcal H^3(SO(3),U(1))$ to topological cohomology $H^3(SO(3),Z)$ is not onto, only
even elements of the latter (namely $K =4r$) have counterparts
of the former \cite{DW9093,GW8693}.

The physical reason that $K$ must be $4r$ is the following.
We consider space-time with $S^1\times \Sigma$ topology,
but in the limit where the spacial circle $S^1$ has a very small size.  Let us consider
the field configuration $g(x^\mu)$ where $S^1$ maps to the nontrivial element
in $\pi_1[SO(3)]=\mathbb Z_2$: $g(x^\mu)=e^{i \theta \hat{\mathbf n}\cdot \mathbf L}$ where
$\theta$ parameterizes the $S^1$ and $L_x,L_y,L_z$ are the generators of the $SO(3)$ group.
In the small $S^1$ limit, such field
configuration is described by the mapping from the space-time $\Sigma$ to $S^2$
labeled by the unit vector $\hat{\mathbf n}$. Physically, this means that the small
$S^1$ limit, $S_{\mathrm{top}}$ can be viewed as the topological $\theta$ term
in the NLSM of unit vector $\hat{\mathbf n}$ with $\theta=2\pi K$, since if $\Sigma$
wrap around $S^2$ once, $g(x^\mu)=e^{i \theta \hat{\mathbf n}\cdot\mathbf L}$ will wrap around
$SO(3)$ twice. In the small $S^1$ limit the space becomes a thin torus (or a cylinder if it is open) and
the system becomes an effective 1D system.  We also note that $g(x^\mu) \to
hg(x^\mu)h^{-1}$, $h\in SO(3)$ rotate the unit vector $\hat{\mathbf n}$.  Such an
$SO(3)$ rotation gives rise to an isospin quantum number $
\mathbf{S}_\text{iso}= \mathbf{S}_{L} +\mathbf{S}_{R} $, where $\mathbf{S}_{L}$
is the spin operator associated with $SO_L(3)$ and $\mathbf{S}_{R}$ with $SO_R(3)$.
The topological $\theta$ term with $\theta=2\pi K$ implies that an open end of
the 1D system will carry isospin $\frac{K}{2}$ \cite{N9455}. This means that a
$Z_2$ vortex (which exists since $\pi_1[SO(3)]=\mathbb Z_2$) will carry
isospin $\frac{K}{2}$.  In Ref. \onlinecite{GW8693}, it is shown that such a
$Z_2$ vortex (corresponding to the twisted sector in Ref. \onlinecite{GW8693})
carries  $(S_{L}, S_{R})$ spins given by $(m+\frac12,\frac{K}{2}-m-\frac12)$,
$m=$ integer, if $K=4r+2$, and by $(m,\frac{K}{2}-m)$, $m=$ integer, if
$K=4r$.  Thus a $Z_2$ vortex carries the physical spin (i.e. the $S_L$ spin)
given by half integers if $K=4r+2$ and by integers if $K=4r$.  $Z_2$ vortex
carrying half-integer spins can happen in the continuous field theory, since
the $Z_2$ vortex is nontrivial in continuous field theory.  However, SPT phases
are defined on lattice models where space-time are discrete. In this case,
the $Z_2$ vortex can continuously deform into a trivial configuration.  Thus
the vortex core must be ``trivial'' and can only carry an integer spin.
Consequently, only $K=4r$ correspond to SPT phases.

Except for the constrains of the level $K=4r$, the remaining discussion is very similar to that of the $SU(2)$ model.
We couple the $SO(3)$ NLSM with an external probe field $A$,
we expect that the effective action for $A$ is a Chern-Simons term (plus a boundary action),
$S_{\mathrm{eff}}(A)= i{K\over16\pi }\int_M {\mathrm{Tr}}(A\wedge F-{1\over3}A^3). $
We can expand $A=\sum_a A^aL_a,\ a=x,y,z$, where $L_x,L_y,L_z$ satisfy $[L_a,L_b]=
i\varepsilon_{abc}L_c$ and $\mathrm{Tr}(L_aL_b)=2\delta_{ab}$. Suppose $A$ is
collinear and only contains the $z$ components in spin space, then we obtain the
response spin current density,
$\mathcal J_\mu^z={\delta S\over\delta A_\mu^z}= i{K\over4\pi}\epsilon^{\mu\nu\lambda}\partial_\nu A_\lambda^z.$
The spin Hall conductance is quantized as ${K\over4\pi}$
[the same as the $SU(2)$ case].

We may embed the edge effective theory into $K/2$ flavor free Majorana fermion model,
\begin{eqnarray*}\label{BoundaryO3}
S_{\mathrm{bdr}}(\psi,A)&=&  \int_{\partial M} dx^0dx^1 \sum_{I=1}^{k}\left[\tilde\psi _{I-} (\partial_\tau-i\partial_\sigma )\tilde\psi_{I-}+\right.\nonumber\\&&\left.{}\tilde\psi _{I+} [(\partial_\tau+  A_\tau)+i(\partial_\sigma+ A_\sigma)]\tilde\psi_{I+}\right],
\end{eqnarray*}
where $\tilde \psi_I$ is a $SO(3)$ triplet Majorana fermion field and $k=K/2$
is the level of $SO(3)$ Kac-Moody algebra. The anomaly of the boundary action
cancels the anomaly of the bulk Chern-Simons term. The field $A$ induces a left
moving spin current on the edge. Again, the extra $O(k)$ gapless modes can be
gapped out by a mass term which does not break the $SO(3)$ symmetry, or can be
removed by a projection onto a $O(k)$ singlet per site.

We may also view the $SO(3)$ SPT phases as $U(1)$ SPT phases.  From the spin
Hall conductance $\frac{K}{4\pi}$ of the $SO(3)$ SPT phases and the fact that
$K=4r$, we see that the $U(1)$ SPT phases have an even-integer quantized Hall
conductance (in units of $1\over2\pi$).

\textit{Conclusion and discussion}. ---
In summary, we study $SU(2)$ and $SO(3)$ symmetry-protected topological phases via topological NLSM. These phases have spin
quantum Hall effect when they are coupled to external probe fields. The gapless
boundary excitations are decoupled left movers and right movers, which
are protected by symmetry. When $K>0$, only the left moving current carries
symmetry charge, and can be detected by the probe field.  The spin Hall
conductance quanta of $SO(3)$ models is 4 times as large as that of the $SU(2)$
models.  We also find that the $U(1)$ SPT phases are characterized by an
even-integer quantized Hall conductance.

It has been shown that different 2D SPT states with symmetry $G$ are described
by Borel group cohomology ${\cal H}^3[G,U(1)]$ \cite{CGL1172}. In this Letter we
show that [for $G=SU(2),SO(3)$] if we  gauge the symmetry group, the
resulting theory is a Chern-Simons theory with gauge group $G$
which is also classified by $\mathcal H^3[G, U(1)]$ \cite{DW9093}.
This suggests a very interesting one-to-one duality relation between 2D SPT phases with
symmetry $G$ and 2D Chern-Simons theory with gauge group $G$, for both
continuous and discrete groups $G$ \cite{LG}. This also suggests that, when we
probe the SPT states by ``gauging'' the symmetry, we can distinguish all the
SPT phases.

We thank Xie Chen, Zheng-Cheng Gu, Liang Fu, Xiong-Jun Liu, Senthil Todadri, and Patrick Lee for helpful
discussions. This work is supported by NSF Grant No. DMR-1005541 and NSFC
11074140. Z.X.L is supported by NSFC 11204149.


\begin{thebibliography}{10}
\bibitem{T} Notice that under transformation $T$, a group element $\hat h$ in $SU(2)_L$ is transferred into its partner $\hat{\bar h}$ in $SU(2)_R$. For instance, $(T\hat hT^{-1})g=T(hg^{-1})=gh^{-1} =\hat{\bar h}g,$ or equivalently $T\hat hT^{-1}=\hat{\bar h}$.
\end{thebibliography}
\end{document}


\newpage

\title{\large \bf  Supplemental material: Symmetry protected Spin Quantum Hall phases in 2-Dimension}

\author{Zheng-Xin Liu}
\affiliation{Institute for Advanced Study, Tsinghua University, Beijing, 100084, P. R. China}
\affiliation{Department of Physics, Massachusetts Institute of Technology, Cambridge, Massachusetts 02139, USA}
\author{Xiao-Gang Wen}
\affiliation{Department of Physics, Massachusetts Institute of Technology, Cambridge, Massachusetts 02139, USA}
\affiliation{Perimeter Institute for Theoretical Physics, Waterloo, Ontario, N2L 2Y5 Canada}

\maketitle

\section{Vectors and tensors in imaginary time}\label{app: time} In this paper,
we used imaginary time formalism. The space-time metric is an identity matrix,
namely, we don't distinguish covariant vector(tensor) or contravariant
vector(tensor). The relation between the imaginary and real time vector is the
following,
\begin{eqnarray}
X^\tau=X_\tau=iX^t=-iX_t.
\end{eqnarray}
For example, the space-time coordinates $r=(\tau,x,y)=(it,x,y)$, derivative $\partial=(\partial_\tau,\partial_x,\partial_y)=(i\partial^t,\partial^x,\partial^y)=(-i\partial_t,\partial_x,\partial_y)$, momentum $k=(k^\tau,k^x,k^y)=(i\omega, k^x,k^y)$, current operators $J=(J_\tau,J_x,J_y)=(iJ^t,J^x,J^y)=(i\rho,J_x,J_y)$, and the external probe field $A=(A_\tau,A_x,A_y)=(iA^t,A^x,A^y)=(-iA_t,A^x,A^y)$.

When deriving formulas, we first work in the imaginary time components, and then map to the real time components using above relations.

\section{relationship between Borel group cohomology and topological cohomology}

The 2+1D SPT phases with symmetry group $G$ discussed in the main text are classified by the Borel group cohomology $\mathcal H^3(G,U(1))$ \cite{CGL1172}. Mathematically, the third order Borel group cohomology group with $U(1)$ coefficients is equal to the forth order topological cohomology group of the classifying space $BG$ with $Z$ coefficients, namely, $\mathcal H^3 (G, U(1))\equiv H^4(BG,Z)$, where $BG$ is the classifying space for fiber bundles with structure group $G$.

For continuous non-Abelian group $G$, the topological theta term of the 2+1D principal chiral nonlinear sigma model is classified by $H^3(G,Z)$. For example, when $G=SU(2)$, $H^3(SU(2),Z)=\mathbb Z$, so there are infinite number of theta terms and hence infinite number of gapped phases accordingly. $H^3(SU(2),Z)$ is generated by the unit volume form on the group manifold: $\omega=\frac{1}{24\pi^2}\int_{G} \mathrm{Tr} (g^{-1}dg)^3$. The topological theta term can be obtained by pulling back the $K$th differential form in $H^3(SU(2),Z)$ onto the space-time manifold $M$,
\begin{eqnarray}\label{Topterm}
S_{\mathrm{top}}=-i\frac{2\pi K}{24\pi^2}\int_{M} \mathrm{Tr} (g^{-1}dg)^3.
\end{eqnarray}

For continuous groups, such as $SU(2)$ or $SO(3)$, there is a map from $H^4(BG,Z)$ to $H^3(G,Z)$, called inverse transgression map. For $G=SU(2)$, this map is one-to-one and onto; while for $G=SO(3)$, this map is not surjective, only half of the elements in $H^3(SO(3),Z)$ have correspondence in $H^4(SO(3),Z)$. Since $\mathcal H^3 (G, U(1))\equiv H^4(BG,Z)$, the inverse transgression map bridges a relation between the Borel group cohomology $\mathcal H^3 (G, U(1))$ and the topological cohomology $H^3(G,Z)$. From above discussion, this map sets up a relationship between the SPT phases (classified by $\mathcal H^3 (G, U(1))$) and principal chiral nonlinear sigma models (classified by $H^3 (G, Z)$). This is the reason that we can study the physical properties of the former from the latter.

Furthermore, in Ref. \cite{DW9093}, it is shown that 2+1D Chern-Simons gauge theory with gauge group $G$ is classified by $H^4(BG,Z)$, no matter $G$ is discrete or continuous. Thus, the classification of 2+1D SPT phase with symmetry group $G$ and the classification of 2+1D Chern-Simons gauge theory with gauge group $G$ is the same. This is the mathematical reason that we can study the response of SPT phases by the corresponding effective Chern-Simons action.

\section {Physical meaning of the $A$ field} \label{app: Afield}

The meaning of $A$ field is a local spin axes twist. For example, if we introduce a new local spin axes such that $\mathbf S'(x)=U(x)\mathbf S(x)$, where $U(x)$ is a $SO(3)$ matrix. Then we obtain a field $A(x)=dUU^{-1}$ (and similarly $\bar A=U^{-1}dU$). This field is nothing but the prob field for the $SO(3)$ NLSM.

We can expand the $A(x)$ field with the spin operators $A(x)=\sum_aA_a(x)S_a$. Then the prob field for $SU(2)$ NLSM is $A(x)$ field is defined as $A_{SU(2)}={1\over2}\sum_a A_a\sigma_a$. The $A$ field defined this way is a `pure gauge', namely, the twisting angle of the spin axes is path independent.

Generally, if the spin field is coupled to an external $SU(2)$ `gauge field', the twist angle of the spin axes maybe path dependent (the Aharonov-Bohm effect). In that case, the strength of the $A$ field is nonzero.

Physically, spin-orbital coupling and lattice distortion (such as phonon) may give rise to $A$.

\section{Relation to the Hopf model}\label{app: hopf}
Now let us go to the topological term
\begin{eqnarray}\label{Topterm}
S_{\mathrm{top}}=-i\frac{\theta}{24\pi^2}\int_{M} \mathrm{Tr} (g^{-1}dg)^3,
\end{eqnarray}
where $g\in SU(2)$. We can introduce three Euler angles to parameterize the $SU(2)$ group,
\begin{eqnarray}
&g=e^{i{\sigma_z\over2}\gamma}e^{i{\sigma_y\over2}\beta}e^{i{\sigma_z\over2}\alpha},\nonumber\\
& 0\leq\alpha<4\pi,   0\leq\beta<\pi,  0\leq\gamma<2\pi,
\end{eqnarray}
then
\begin{eqnarray}
g^{-1}dg&=&{i\over2}\left[(\sin\beta\cos\alpha\sigma_x+\sin\beta\sin\alpha\sigma_y+\cos\beta\sigma_z)d\gamma\right.\nonumber\\&&\left.
+(\cos\alpha\sigma_y-\sin\alpha\sigma_x)d\beta+\sigma_zd\alpha\right]
\end{eqnarray}
and
\begin{eqnarray}
\mathrm{Tr} (g^{-1}dg)^3={3\over2}\sin\beta d\beta\wedge d\alpha\wedge d\gamma.
\end{eqnarray}
Under `adiabatic' condition, we can introduce the `Berry connection', which is the diagonal part of $g^{-1}dg$,
\begin{eqnarray}
a={1\over2}\mathrm{Tr}(\sigma_z g^{-1}dg)\sigma_z={i\over2}\sigma_z(\cos\beta d\gamma + d\alpha),
\end{eqnarray}
Then the Berry curvature reads $f=da=-{i\over2}\sigma_z\sin\beta d\beta\wedge d\gamma$, and the topological term (\ref{Topterm}) can be identical to the hopf term,
\begin{eqnarray}
S_{\mathrm{top}}=-i\frac{K}{4\pi}\int_M dx^3\mathrm{Tr}(a\wedge f),
\end{eqnarray}
Notice that our model is a little bit different from that discussed by Wilczek and Zee \cite{WilZee}, because our $\theta$ is quantized.

Suppose the external field $A$ acts on group elements in the same way as the internal Berry connection $a$, then we expect that the effective field theory should be
\begin{eqnarray}\label{Aa}
S_{\mathrm{top}}=-i\frac{K}{4\pi}\int_M dx^3\mathrm{Tr}[(a+A)\wedge (f+F)],
\end{eqnarray}
here we have assumed that $A$ and $F$ only have $\sigma_z$ component. When integrating out the group variables, we obtain an effective action $S_{\mathrm{eff}}(A)$. Comparing Eq.~ (\ref{Aa}) and Eq.~(3) in the main text, the $A\wedge F$ part is present in the effective action here, but the $A^3$ term is missing. We guess that a gauge invariant effective action takes the following form
\begin{eqnarray}
S_{\mathrm{eff}}(A)= i{K\over4\pi }\int_M \mathrm{Tr}(A\wedge F-{1\over3}A^3) + S_{\mathrm{bdr}}(A).
\end{eqnarray}

\section{derivation of the gauge anomaly for chiral fermions}\label{app: cal}
We introduce the 1+1D gamma matrices
\[\gamma_0=\sigma_x,\ \gamma_1=-\sigma_y, \ \gamma_5=i\gamma_0\gamma_1=  \sigma_z,\]
which satisfy $\{\gamma_\mu,\gamma_\nu\}=0$ for $\mu\neq\nu$ and $(\gamma_0)^2=(\gamma_1)^2=(\gamma_5)^2=1$. With these matrices, we can combine the right moving and left moving fermions as Dirac fermions $\psi=(\psi_-,\psi_+)^T$ and $\bar\psi=\psi^\dag\gamma_0$, where $\psi_\mp={1\pm \gamma_5\over2}\psi$ is the eigenstates of $\gamma_5$ with eigenvalues $\pm1$. The total action for the fermions on the boundary can be written as
\begin{eqnarray}\label{FermBdr}
S_{\mathrm{bdr}}(\psi_I,A)&=&  \int_{\partial M} dx^0dx^1 \left[\psi^\dag_{I-} (\partial_\tau-i\partial_\sigma )\psi_{I-}\right.\nonumber\\&&\left.{}+\psi^\dag_{I+} [(\partial_\tau+{}A_\tau) +i(\partial_\sigma+A_\sigma)]\psi_{I+}\right]\nonumber\\
&=& \int_{\partial M} d^2x\left[\bar\psi_I\slashed\partial{1+\gamma_5\over2}\psi_I+\bar\psi_I\slashed D{1-\gamma_5\over2}\psi_I \right].\nonumber\\
\end{eqnarray}
where $\slashed D=\slashed \partial+\slashed A$, with $\slashed \partial=\sum_\mu\gamma^\mu\partial_\mu$, and $\slashed A=\sum_\mu\gamma^\mu A_\mu$. Notice that the gauge field only couple to the right-moving chiral fermions.

We introduce the gauge transformation $e^{iv(x){1-\gamma_5\over2}}$ for $\psi$. It is equivalent to gauge transform the right moving fermions $\psi_+$ by $h=e^{iv(x)}$. Noting ${1-\gamma_5\over2}\gamma_0=\gamma_0{1+\gamma_5\over2}$, we have
\begin{eqnarray}\label{gaugetr}
\psi_+'&=&h\psi_+,\ \ \bar\psi_-'=\bar\psi_-h^{-1},\nonumber\\
A'&=&hAh^{-1}+hdh^{-1}.
\end{eqnarray}
For convenience, we can expand $hd h^{-1}=-idV(x)$ where $V(x)$ is determined by $v(x)$. If $v(x)$ is small, then $V(x)\approx v(x)$.

We can calculate the gauge anomaly using path integral method following Fujikawa. Since the calculation for different flavors are the same, we will only consider a single flavor and omit the subscript $I$. According to Eq.~(\ref{FermBdr}), the effective action was given as
\begin{eqnarray}\label{path}
&&e^{-S_{\mathrm{bdreff}}(A)}=\int \mathcal D\bar\psi \mathcal D\psi e^{-\int_{\partial M} d^2x \bar\psi \hat D\psi},
\end{eqnarray}
where $\hat D=\slashed D{1-\gamma_5\over2}+\slashed \partial{1+\gamma_5\over2}$. So we can expand $\psi$ by the eigenstates of $\hat D$, $\psi=\sum_n a_n\phi_n$ and $\bar\psi= \sum_n \bar b_n\chi_n^\dag$, with $\hat D\phi_n=\lambda_n\phi_n,\ \hat D^\dag\chi_n=\lambda_n^*\chi_n$ and $\int \chi_n^\dag\phi_m=\delta_{mn}$. Then $\mathcal D\bar\psi \mathcal D\psi=\mathcal D\bar b\mathcal Da=\prod_nd\bar b_nda_n$ and $\int d^2x\bar\psi \hat D\psi = \sum_n\lambda_n\bar b_na_n$. The integral in Eq.~(\ref{path}) is formally equal to $\det (\hat D)=\prod_n\lambda_n$.

Now we consider the gauge transformation $e^{iv{1-\gamma_5\over2}}$ for $\psi$. When $v$ is small, we can keep the leading order terms of $v$ in Eq.~(\ref{gaugetr})
\begin{eqnarray}\label{gauge2}
\psi'&\approx&\psi+iv{1-\gamma_5\over2}\psi,\nonumber\\
\bar\psi'&\approx&\bar\psi-i\bar\psi{1+\gamma_5\over2}v
\nonumber\\A_\mu'&\approx& A_\mu-i\partial_\mu v-i[A_\mu, v].
\end{eqnarray}
It can be shown that the modulus $|\det (\hat D)|$ is invariant under the gauge transformation.\cite{ZWZ8477,AG8449} The anomaly comes from the measure when changing the variables in Eq.~(\ref{gauge2}),
\begin{eqnarray}\label{J1J2}
\mathcal D\bar\psi' \mathcal D\psi'=\mathcal D\bar b' \mathcal Da'=(J_1J_2)^{-1}\mathcal D\bar b \mathcal Da,
\end{eqnarray}
$J_1$($J_2$) is the Jacobian of $\bar b$($a$),
\begin{eqnarray}
\ln(J_1J_2)&=&\ln[\det(1+C_1)\det(1+C_2)]\nonumber\\&=&\mathrm{Tr}[\ln(1+C_1+C_2+...)]\nonumber\\
&\approx&\mathrm{Tr}(C_1+C_2),
\end{eqnarray}
where $\bar b'=\bar b(1+C_1)$ with $(C_1)_{mn}=\int_{\partial M} d^2x (-i v\chi_m^\dag{1+\gamma_5\over2}\phi_n)$ and $a'=(1+C_2)a$ with $(C_2)_{mn}=\int_{\partial M} d^2x (i v\chi_m^\dag{1-\gamma_5\over2}\phi_n)$. The traces $\mathrm{Tr}(C_1+C_2)$ is divergent. We can introduce a cut off $M_0$ to regularize it in momentum space. Noticing that $\hat D^2=\sls D\sls \partial{1+\gamma_5\over2}+\sls \partial\sls D{1-\gamma_5\over2}$, 
we have
\begin{eqnarray}
\ln(J_1J_2)&=&\int_{\partial M} d^2x(-iv)\sum_n\chi_n^\dag\gamma_5\phi_ne^{{\lambda_n^2\over M_0^2}}|_{M_0\to\infty}\nonumber\\
&=&-\int d^2xiv\sum_n\chi_n^\dag(x)\gamma_5e^{{\hat D^2\over M_0^2}}\phi_n(y)|_{y\to x,M_0\to\infty}\nonumber\\
&=&-\int d^2 x iv\mathrm{Tr}\left[{\gamma_5\over2M_0^2}[\gamma_\mu,\gamma_\nu]\partial_\mu (A'_\nu)\right]\nonumber\\&&\times\int {d^2k\over(2\pi)^2}e^{-{k^2\over M_0^2}}|_{M_0\to\infty}\nonumber\\
&=&{i\over4\pi}\int_{\partial M}\mathrm{Tr}(ivdA)\nonumber\\
&\approx&-{i\over4\pi}\int_{\partial M}\mathrm{Tr}(h^{-1}dh\wedge A).
\end{eqnarray}
Here we only kept the linear term with $v(x)$. From Eqs.~(\ref{path}) and (\ref{J1J2}), considering there are $K$ flavors of fermions, we have $\delta S_\mathrm{bdreff}(A)=K\ln(J_1J_2)=-{iK\over4\pi}\int_{\partial M}\mathrm{Tr}(h^{-1}dh\wedge A)$, which cancels the anomaly of Eq.~(5) in the main text.




%